\documentclass{jfm}
\usepackage{graphicx}
\usepackage[dvipsnames]{xcolor}
\usepackage{amsmath}
\usepackage{amssymb}

\shorttitle{Near-resonant instability of geostrophic modes }
\shortauthor{T. Le Reun, B. Gallet B. Favier and M. Le Bars}

\title{Near-resonant instability of geostrophic modes:  beyond Greenspan's theorem}

\author{T. Le Reun \aff{1,2} \corresp{\email{tl402@cam.ac.uk}},
  B. Gallet \aff{3},
  B. Favier \aff{1}
 \and M. Le Bars\aff{1}}

\affiliation{\aff{1} Aix Marseille Univ, CNRS, Centrale Marseille, IRPHE UMR 7342, Marseille, France
\aff{2} DAMTP, University of Cambridge, Wilberforce Road, Cambridge CB3 0WA, UK
\aff{3} Service de Physique de l'\'Etat Condens\'e, CEA Saclay, CNRS, Universit\'e Paris-Saclay
}
 \definecolor{brickred}{rgb}{0.8, 0.25, 0.33}
\xdefinecolor{RED}{named}{black}
\xdefinecolor{RED2}{named}{brickred}

\graphicspath{{./images/}}

\newcommand{\mbf}{\boldsymbol}
\newcommand{\mathscr}{\mathcal}

\newcommand{\ie}{\textit{i.e.}~}
\newcommand{\bu}{\mbf{u}}

\newcommand{\bk}{\mbf{k}}
\newcommand{\bq}{\mbf{q}}
\newcommand{\bp}{\mbf{p}}
\newcommand{\bU}{\mbf{U}}

\newcommand{\ddroit}[2]{\frac{\mathrm{d} #1}{\mathrm{d} #2}}

\newcommand{\bQ}{\mbf{Q}}
\newcommand{\bK}{\mbf{K}}

\newcommand{\norm}[1]{\left\lVert#1\right\rVert}

\begin{document}
\xdefinecolor{RED2}{named}{black}

\maketitle 
\begin{abstract}
We explore the near-resonant interaction of inertial waves with geostrophic modes in rotating fluids via numerical and theoretical analysis.
When a single inertial wave is imposed, we find that some geostrophic modes are unstable \textcolor{RED2}{above a threshold value of the Rossby number $kRo$ based on the wavenumber and wave amplitude.}
We show this instability to be caused by triadic interaction involving two inertial waves and a geostrophic mode such that the sum of their eigen frequencies is non-zero. 
We derive theoretical scalings for the growth rate of this near-resonant instability.
\textcolor{RED2}{
The growth rate scaled by the global rotation rate is proportional to $(kRo)^2$ at low $kRo$ and transitions to a $kRo$ scaling for larger $kRo$. 
}
These scalings are in excellent agreement with direct numerical simulations.
This instability could explain recent experimental observations of geostrophic instability driven by waves. 
\end{abstract}

\section{Introduction}

Rotating turbulent flows are ubiquitous in geo- and astrophysical systems such as stellar interiors, planetary cores, oceans and atmospheres. 
In a large number of numerical simulations and experiments (see the review by \cite{godeferd_structure_2015}), rotating turbulence is observed to develop a strong anisotropy and to spontaneously form vortices that are invariant along the rotation axis.
The latter correspond to a first-order balance between the Coriolis force \textcolor{RED2}{and pressure gradients} and are called ``geostrophic modes''. 
\textcolor{RED}{
Yet, the systematic observation of strong geostrophic modes is at odds with various evidence suggesting that rotating turbulence could as well be dominated by inertial waves that are sustained by the restoring action of the Coriolis force.}
Recent numerical \citep{le_reun_inertial_2017} and experimental \citep{le_reun_experimental_2019,brunet_shortcut_2020} studies have shown that injecting energy in waves solely creates a turbulent state comprising of inertial waves only when the forcing amplitude is sufficiently small, \ie a discrete version of inertial wave turbulence \citep{galtier_weak_2003}. 
It is only at larger forcing amplitudes that a secondary instability leads to the classical geostrophic-dominated turbulence.
\textcolor{RED}{Asymptotic theories describing rotating turbulence in the limit  of vanishing forcing amplitude and dissipation also suggest that waves could dominate over geostrophic modes in such a regime \citep{bellet_wave_2006,sagaut_homogeneous_2018}.}
Hence, although bi-dimensionalisation in the form of geostrophic eddies has been commonly observed, it may not be the only equilibrium state of rotating turbulence, \textcolor{RED}{be it at moderate \citep{yokoyama_hysteretic_2017-1,favier_subcritical_2019} 
as well as small \citep{van_kan_critical_2019} forcing amplitudes.}
\textcolor{RED}{In addition, the nature of the forcing seems fundamental in determining the equilibrium state of rotating turbulence}. 
\textcolor{RED}{These results altogether call for a better understanding of the fundamental processes by which waves give rise to balanced geostrophic modes.} 
%
%The emergence of balanced geostrophic flows out of waves remains a challenging problem.
%
 \textcolor{RED}{The studies of \cite{le_reun_experimental_2019} and \cite{brunet_shortcut_2020}
suggest that such a transfer occurs through an instability.}
Although wave-to-wave interactions are primarily governed by triadic resonance \citep{bordes_experimental_2012,
vanneste_wave_2005}, they cannot account for wave-to-geostrophic transfers \citep{greenspan_non-linear_1969}, at least in the asymptotic limit of vanishing velocity amplitude and dissipation.
Several alternative mechanisms, outside the framework of Greenspan's theorem, have been proposed. 
Four-modes interactions can transfer energy from waves to geostrophic flows, either directly \citep{newell_rossby_1969,smith_transfer_1999} or through an instability mechanism \citep{kerswell_secondary_1999,brunet_shortcut_2020}.
The growth rate of such an instability scales like $Ro^2$, with $Ro$ the dimensionless wave amplitude or Rossby number.
It develops over longer timescales that triad-type interactions between waves.
The other inviscid mechanism that has been proposed to account for wave-geostrophic transfer is quasi-resonant triadic interaction \citep{newell_rossby_1969,smith_transfer_1999}, that is, a triad between waves whose frequencies do not exactly satisfy the resonance condition \citep{bretherton_resonant_1964}.
Their presence and their role in the bi-dimensionalisation of rotating turbulence has been assessed by several numerical studies \citep{smith_near_2005,alexakis_rotating_2015,
clark_di_leoni_quantifying_2016}.
While it has been shown that such triads can transfer directly energy from two pre-existing waves to geostrophic modes, we show that this transfer can arise spontaneously through an instability mechanism.
\textcolor{RED}{More precisely, we show with direct numerical simulations (DNS) and theoretical analysis that there exists a linear mechanism by which a single inertial wave drives exponential growth of geostrophic modes through near-resonant triadic interaction.
}

\section{The stability of a single inertial wave}
\label{sec:DNS_section}

\subsection{Governing equations and numerical methods} 

Let us consider an incompressible fluid rotating at rate $\Omega \mbf{e}_z$. 
We investigate the stability of a single plane inertial wave  with wave vector $\bk$ and eigen frequency $\omega_k$. 
Its amplitude is proportional to $\bU_w$ with
\begin{equation}
\label{eq:ch2_plane_wave}
\mbf{U}_w = \mbf{h}_{\mbf{k}}^{s_k} \exp i(\mbf{k}\cdot\mbf{x}- \omega_k t )~ + c.c.
\end{equation}
$\mbf{h}_{\mbf{k}}^{s_k}$ is a helical mode, that is, when $\bk$ is not parallel to the axis of rotation $\mbf{e}_z$ \textcolor{RED}{\citep{cambon_spectral_1989,waleffe_nature_1992}}
\begin{equation}
\mbf{h}_{\bk}^{s_k} = \frac{1}{\sqrt{2}} \left(\frac{ \left(\bk \times \mbf{e}_z \right) \times \bk}{\vert  \left(\bk \times \mbf{e}_z \right) \times \bk \vert }  + i s_k  \frac{\bk \times \mbf{e}_z}{\vert \bk \times \mbf{e}_z \vert} \right) ,
\end{equation} 
where $s_k = \pm 1$ is the sign of the helicity of the plane wave.
If $\bk$ is parallel to $ \mbf{e}_z$, $\mbf{h}_{\mbf{k}} = \mbf{e}_x + i s_k  \mbf{e}_y$. 
$\bU_w$ automatically satisfies the incompressibility condition since $ \bnabla \cdot \bU_w$ $ \propto \bk \cdot \mbf{h}_{\bk}^{s_k} = 0 $.
$\bU_w$ satisfies the linearised rotating Euler equation,
\begin{equation}
\partial_t \bU_w + 2 \mbf{\Omega} \times \bU_w = - \mbf{\nabla} \pi_w ~,
\end{equation}
provided that $\omega_k^{s_k}$ and $\bk$ are related by the dispersion relation of inertial waves
\begin{equation}
\label{eq:ch2_reldisp}
\omega_k = 2 s_k \Omega \frac{k_z}{k} = 2 s_k \Omega \cos \theta,
\end{equation}
$\theta$ being the angle between the wavevector $\bk$ and the rotation axis $\mbf{e}_z$ (ranging from $0$ to $\pi$), and $k = \vert \bk \vert$. 
We solve for the time evolution of perturbations $\bu$ to the wave $\bU_w$ maintained at a constant amplitude via the following set of equations 
 \begin{subequations}
 \label{eq:ch2_full_problem_equation}
\begin{align}
\partial_t \bu + Ro \left(\mbf{U}_w \cdot \mbf{\nabla} \bu +\bu \cdot \mbf{\nabla} \mbf{U}_w\right) + \bu \cdot \mbf{\nabla} \bu + 2 \mbf{e}_z \times \bu& =
- \mbf{\nabla} \pi + E \mbf{\nabla}^2 \bu  \\
\bnabla \cdot \bu &= 0
\end{align}
 \end{subequations}
where time is scaled by $\Omega^{-1}$ and length by the domain size $L$.
%\footnote{This length will be explicited hereafter.}.
%
We have introduced the Ekman number $E = \nu /(L^2 \Omega)$, $\nu$ being the kinematic viscosity, and an input Rossby number $Ro$ quantifying the dimensionless amplitude of the plane wave.

%\subsection{Numerical methods}
%

%
Equations (\ref{eq:ch2_full_problem_equation}) are solved numerically in a triply periodic cubic box %of size $L = 1$
using the code Snoopy \citep{lesur_relevance_2005}. 
The dynamics of the perturbation flow $\left\lbrace \bu, \pi \right\rbrace$ is determined with pseudo-spectral methods, that is, $\left\lbrace \bu, \pi \right\rbrace$ is decomposed into a truncated sum of Fourier modes $\left\lbrace \hat{\bu}_{\bq},\hat{\pi}_{\bq}  \right\rbrace e^{ i \bq \cdot \mbf{x}} $.
A wave-vector $\bq$ writes $ 2 \pi/L \left( n_x, n_y, n_z \right)$ where \textcolor{RED}{$n_{x,y}$ are integers varying from $-N$ to $N$ and $n_z$ is an integer ranging from $0$ to $N$ because of Hermitian symmetry.} 
In the following, $N = 96$; higher resolutions have been tested and yield the exact same results.
The temporal dynamics of the modes $\hat{\bu}_{\bq}$ is solved using a third order Runge-Kutta method. 
%
%The interaction between the maintained wave $\bU_{w}$ and the perturbation $\bu$ is implemented directly in the non-linear term. 
%%
Note that the size of the box $L$ is artificial and we thus expect our results to depend on the intrinsic Rossby number based on the imposed wavelength $k Ro$ rather than $Ro$ alone. 

%
%Setting $\bU_w$ as an initial condition instead leads to very similar results provided that the Ekman number is sufficiently small to prevent the viscous decay of the imposed wave. 

%
\subsection{Numerical results}
\begin{figure}
\centering
\includegraphics[width=0.496\linewidth]{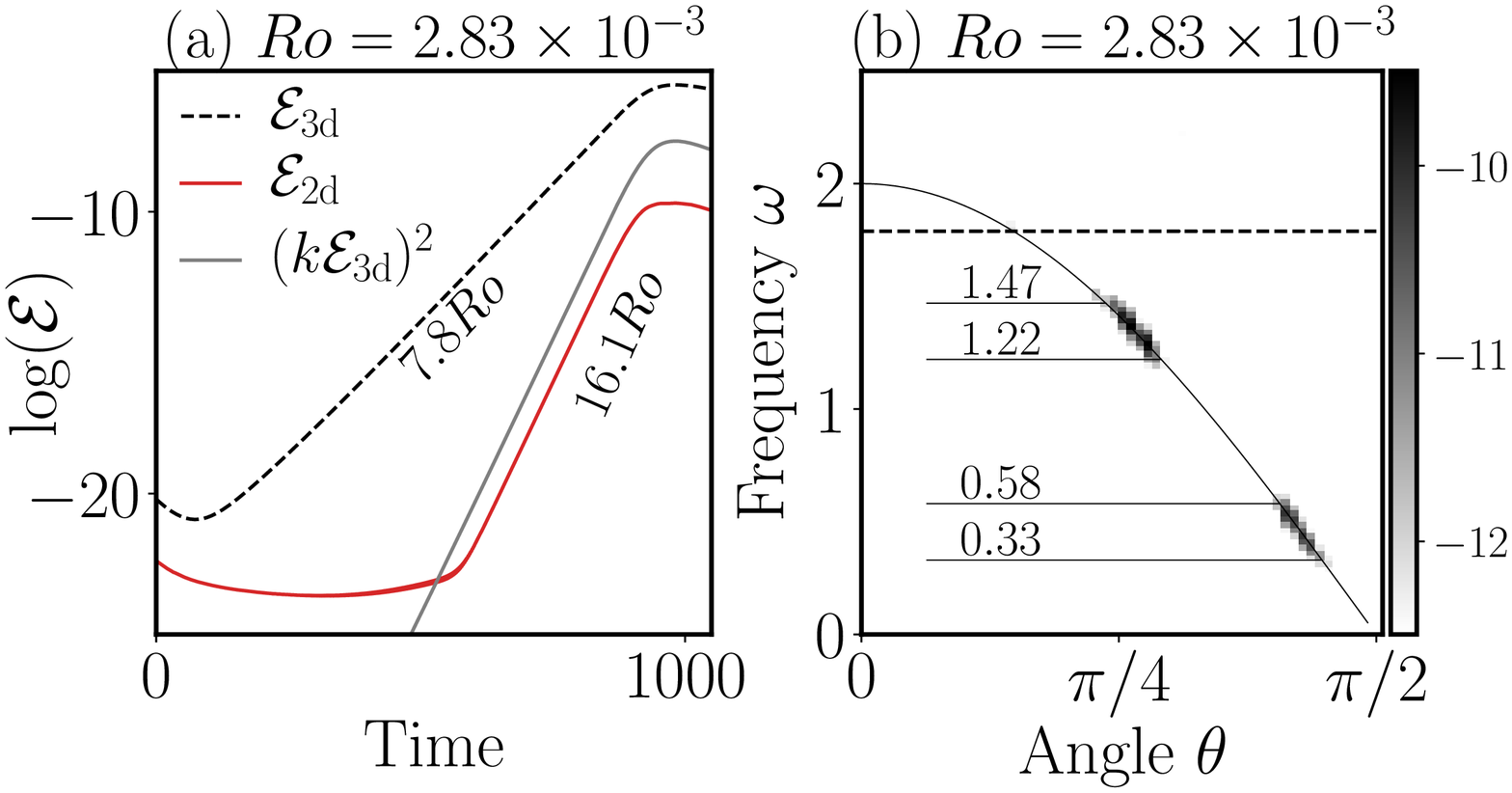}
\includegraphics[width=0.496\linewidth]{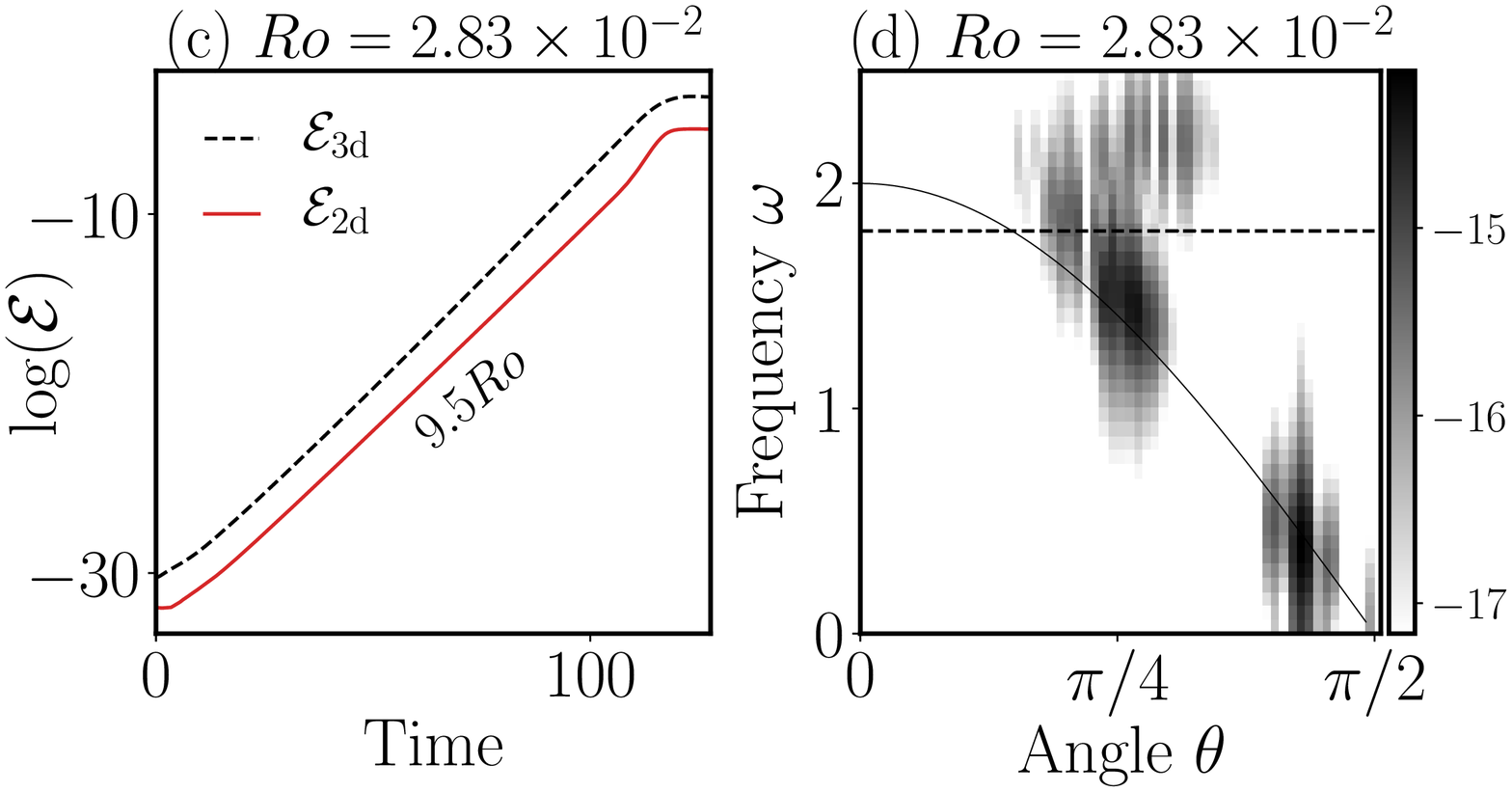}
\caption{Kinetic energy time series (a and c) and heat map of  $\log(\mathcal{E}(\theta,\omega))$ (b and d) resulting from two numerical simulations of the stability of the inertial wave $\bk = 2 \pi \left[4,0,8 \right]$ at $Ro = 2.83 \times  10^{-3}$ and $Ro= 2.83 \times 10^{-2}$. 
In panels (a) and (c), the labels indicate the slope of the best fit for the exponential growth. 
In panel (b) and (d), the plain line materialises the dispersion relation of inertial waves and the horizontal dashed line the frequency of the imposed wave ($ \omega_k \simeq 1.78$).
\textcolor{RED}{For the spectral energy maps, the temporal Fourier transforms have been performed until $t= 800$ for panel $(b)$ and $t= 100$ for panel (d). }
In panel (b), we have indicated the extremal frequencies of the two energy locations. 
}
\label{fig:amplitude_kinetic_map}
\end{figure}

Keeping the Ekman number to $E = 10^{-6}$, two simulations of the stability of the inertial wave $\bk = 2 \pi \left[ 4,0,8 \right]$ (with $s_k = 1$) are performed at low ($Ro = 2.83 \times 10^{-3}$) and moderate ($Ro = 2.83 \times 10^{-2}$) wave amplitudes. 
They are both initiated with a random noise comprising \textcolor{RED}{wavenumbers} ranging between $0$ and $15 \pi$, with very small initial amplitude.
The use of spectral methods allows separating the kinetic energy of the  perturbation flow $\bu$ into a two-dimensional component \textcolor{RED}{$\mathcal{E}_{\rm{2d}}$}, accounting for all modes $\bq$ with $q_z=0$, and a complementary three-dimensional component $\mathcal{E}_{\rm{3d}}$. 
In addition, performing Fourier transform in space and time allows projecting the kinetic energy of $\bu$ in the sub-space of the dispersion relation of inertial waves to draw spatio-temporal spectrum $\mathcal{E}(\theta, \omega)$ \citep{yarom_experimental_2014-1,le_reun_inertial_2017}.
Note that, in these maps, $\theta$ is restricted to $\left[0, \pi/2\right]$ since the flow is real and only the wavevectors $\bq$ with $q_z \geq 0$ are simulated due to the Hermitian symmetry.
\textcolor{RED}{Moreover, the spectra are symmetrised with respect to $\omega=0$  and the maps are shown as a function of $ \vert \omega\vert $ to be more compact.}

\begin{figure}
\centering
\includegraphics[width=0.43\linewidth]{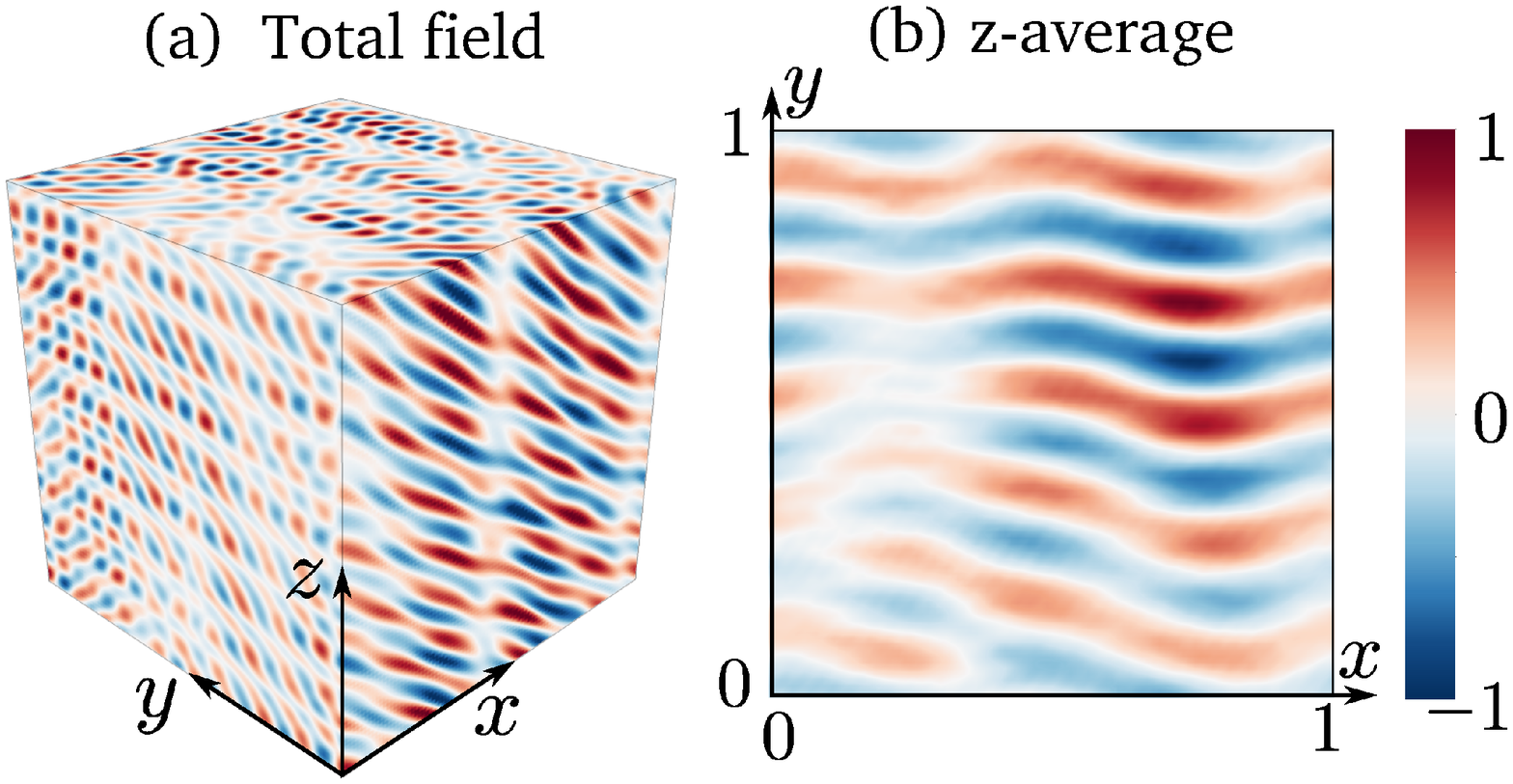}
\includegraphics[width=0.56\linewidth]{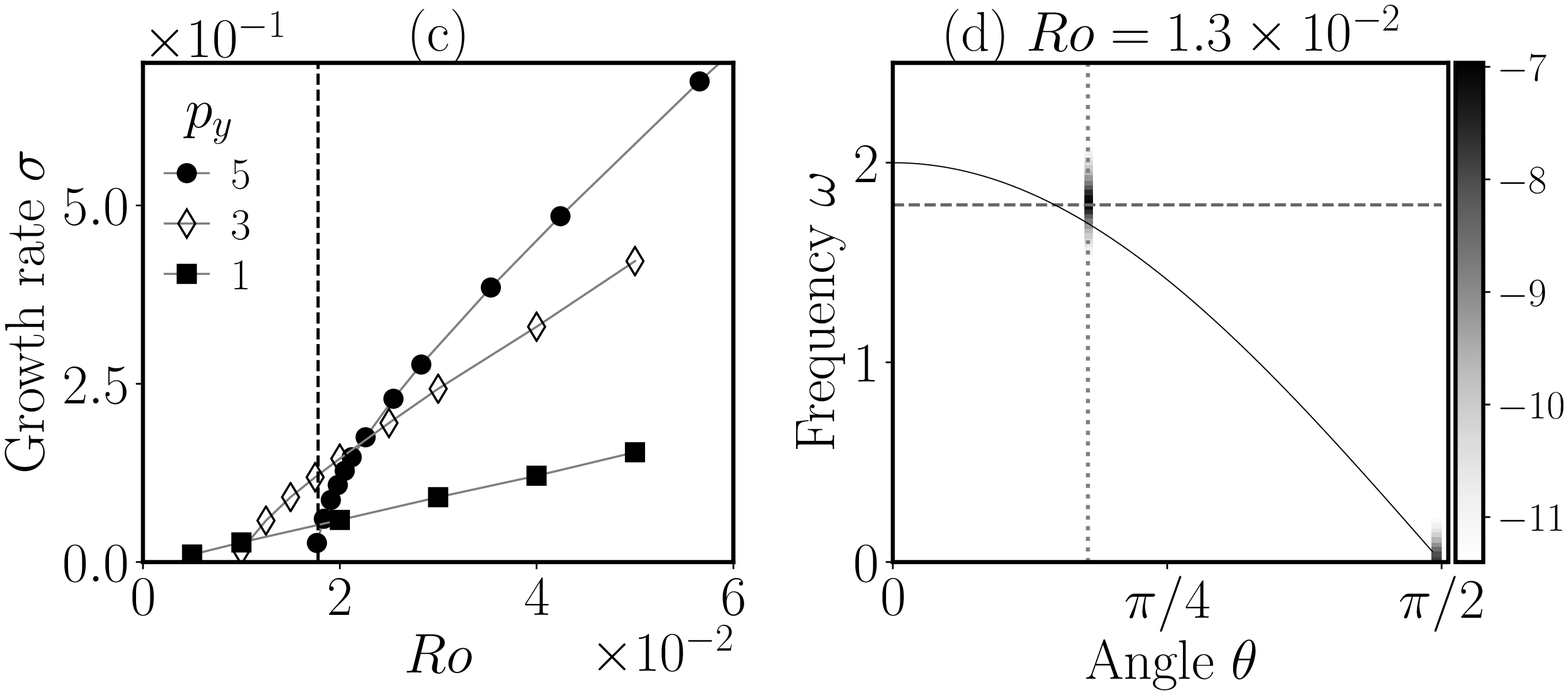}
\caption{
(a) Three-dimensional vertical vorticity field of the growing perturbation at $Ro = 2.83 \times 10^{-2}$ and (b) its vertical average. The fields are normalised by their maximum values.
(c) Growth rate of the geostrophic modes $\bp_g = 2 \pi \left[0, p_{y}, 0 \right]$ with $p_{y} \in \left\lbrace 1,3,5 \right\rbrace$ as a function of the imposed wave amplitude $Ro$. 
The vertical line materialises $k Ro = 1$. 
\textcolor{RED}{The lines joining the markers are used to facilitate the identification of each curve.}
(d) Heat map of $\log(\mathcal{E}(\theta,\omega))$ in the case $p_{y} = 3$ and $Ro = 1.3 \times 10^{-2}$. The same lines as in figures \ref{fig:amplitude_kinetic_map} (b and d) are reported, and a vertical line materialises the angle of the modes closing the triad, that is $-\bk \pm \bp$.}
\label{fig:growth_rate_Ro}
\end{figure}

The kinetic energy time series and maps are shown for both simulations in figure \ref{fig:amplitude_kinetic_map}. 
At low imposed wave amplitude (figure \ref{fig:amplitude_kinetic_map} a-b), three-dimensional perturbations dominate the growth of the instability. 
The energy map displays two spots aligned on the dispersion relation with negative frequencies $\omega_{1,2}
$ such that $\omega_k + \omega_1 +\omega_2 = 0$  which is indicative of several waves undergoing triadic resonance with the imposed mode.
The growth of two-dimensional modes is delayed and their growth rate is approximately twice larger than the rate of three-dimensional modes. 
\textcolor{RED}{Removing the non-linear term $\bu \cdot \bnabla \bu$ (see equation (\ref{eq:ch2_full_problem_equation})) suppresses the growth of two-dimensional modes which are thus} not unstable themselves, at least on the timescale of the growth of waves.
\textcolor{RED}{
In fact, we find that $\mathcal{E}_{\mathrm{2d}} \simeq 3 \times 10^{-2} (k \mathcal{E}_{\mathrm{3d}})^2$ (see figure \ref{fig:amplitude_kinetic_map}a) in the growth phase which suggests that two-dimensional modes' growth} is due to direct forcing by non-linear interaction of two growing waves involved in triadic resonances, with close frequencies and opposed vertical wavenumbers.
This mechanism corresponds to the direct excitation of geostrophic modes by two waves identified by \cite{newell_rossby_1969} and \cite{smith_transfer_1999}.
At larger wave amplitude (figure \ref{fig:amplitude_kinetic_map}c-d), this picture changes: two- and three-dimensional modes grow at the same rate from the start of the simulation.
The vertical vorticity field of the growing perturbation and its vertical average are displayed in figure \ref{fig:growth_rate_Ro}a-b.
\textcolor{RED}{The frequency of two-dimensional modes ($\theta =\pi/2$) is close to $0$ according to figure \ref{fig:amplitude_kinetic_map}d, they are thus geostrophic, that is, slow as well as invariant along the $z$ axis.}
The geostrophic unstable flow contains the wavevector $\bp_g = 2 \pi \left[0,5,0 \right]$ (and $-\bp_g$ for hermiticity), which grows along with other three-dimensional structures whose energy location in the $(\theta, \omega)$ space is reminiscent of triadic resonant instability, but with significantly more spreading. 
\textcolor{RED}{Note that the properties of the transition and the instability we report are robust to changes in the aspect ratio of the box, which discards any spurious effect of the discretisation \citep{smith_near_2005}.}

To further characterise the growth of geostrophic modes, we carry out simulations with the same imposed wave, but we use as initial condition only one wavevector, $\bp = 2 \pi \left[0,p_{y},0 \right]$ with $p_{y} \in \left\lbrace 1,3,5 \right\rbrace$ and $s_{p} = \pm 1$, instead of random noise.
The Ekman number is set to $10^{-8}$ to discard effects of viscosity in the mode growth. 
The growth rate of geostrophic modes is reported in figure \ref{fig:growth_rate_Ro}c where $Ro$ is systematically varied. 
At large wave amplitude, the growth rate increases linearly, but it goes to zero at a finite value of $Ro \sim 10^{-2}$.
\textcolor{RED}{This value is very large compared to the viscous damping $k^2 E \sim 10^{-5}$ which plays no role here,} hence suggesting an inviscid mechanism.
On the kinetic energy map $\mathcal{E}(\theta,\omega)$ (figure \ref{fig:growth_rate_Ro}d) two particular spots appear, one at the location of geostrophic modes and the other at the angle of the vectors closing the triads between $\bk$ and $\pm \bp_g$, which is indicative of triad-type interactions.
This may seem \textit{a priori} in contradiction with the result of \cite{greenspan_non-linear_1969}.
Yet, our results are outside Greenspan's theorem framework in two aspects.
First, our study is necessarily limited to finite Rossby number.
Second, the frequency of the closing mode $-\bk \pm \bp$ (figure \ref{fig:growth_rate_Ro}d) is outside its eigen frequency.
\textcolor{RED}{
In fact, the discrepancy between the observed and eigen frequencies of the closing mode corresponds to the sum of the eigen frequencies of the three modes, $\Delta \omega_{kpq}$.
Moreover, as $Ro$ is decreased, the fastest growing mode $\bp_g$ in figure \ref{fig:growth_rate_Ro}c goes from $p_y = 5$ with $\Delta \omega_{kpq}=0.22$ to $p_y = 1$ with $\Delta \omega_{kpq}=0.01$: the triad is nearly resonant and draws closer to exact resonance
Two mechanisms may be considered to explain the growth of geostrophic modes.
%
%On the one hand, the results are compatible with a near-resonant triadic instability.
%
One hypothesis is that the growth of the geostrophic modes is a by-product of a classical resonance between, say, $\bk$ and $-\bq$ that forces $\bq$ (by Hermitian symmetry) and hence $\bp$. 
However, in this case, the mode closing the wave triad would have an eigen frequency $-\omega_k + \omega_q = \Delta \omega_{kqp} -2 \omega_k$ which is outside the range of inertial waves for small $\Delta \omega_{kpq}$ since $\omega_k \simeq 1.78$.
Instead, our results point towards a near-resonant triadic instability that transfers energy from an inertial wave to a $z$-invariant geostrophic mode.
}
\section{Near-resonance of geostrophic modes: theoretical approach}

\subsection{The low Rossby number limit}
\label{sec:single_triad}

%\textcolor{RED}{In this section, we use asymptotic expansion to derive the growth rate of the near-resonant instability of geostrophic modes in the low Rossby number limit.
%%
%Using plane wave decomposition, we study the evolution of slowly varying amplitudes of modes in a near-resonant triad involving an imposed inertial wave and a $z$-invariant geotrophic mode.
%%
%We show that it is only when the detuning between the eigen frequencies is non-zero but $O((kRo)^2)$ at most that the imposed wave is able to the geostrophic mode.
%%
%By expanding the dynamics of the triad in the neighbourhood of exact resonance, we find the maximum growth rate to scale like $(k Ro)^2$ (equation (\ref{eq:Ro2_scaling})).
%%
%}
\textcolor{RED2}{To provide theoretical insight into the inertial-wave destabilisation observed in the numerical simulations, we turn to linear stability analysis. 
The flow is $\bU = \bu + Ro\, \bU_w$ where  $Ro \,\bU_w$ is the base flow ($Ro$ being the finite Rossby number) and $\bu \ll Ro\, \bU_w$ is now an infinitesimal perturbation.
We consider a Cartesian domain with periodic boundary conditions and we decompose $\bU$} into a superposition of plane waves $\mbf{h}_{p}^{s_p} \exp i (\bp \cdot \mbf{x}-\omega_{p}^{s_p} t)$ 
with time-dependent amplitudes $b_{p}^{s_p} (t)$.
The Euler equation governing $\bU$ is then equivalent to a set of ordinary differential equations governing the amplitudes $b_{p}^{s_p}$ \citep{smith_transfer_1999}:
\begin{equation}
\label{eq:EUleur_ODE}
\ddroit{b_{p}^{s_p}}{t}= \sum_{\bk + \bp + \bq = \mbf{0}}  \sum_{s_k, s_q = \pm 1 } C_{pkq}^{s_p s_k s_q}\, b_{k}^{s_k*}\, b_{q}^{s_q*}\, \exp \left( i \Delta \omega_{k p q} t \right) 
\end{equation}
\begin{equation}
\label{eq:detuning_coupling_definitions}
\mbox{with}~~~ C_{pkq}^{s_p s_k s_q} \equiv \frac{1}{2} (s_{q} q - s_k k) \, \mbf{h}_{p}^{s_p\,*} \cdot \left(\mbf{h}_{k}^{s_k *} \times \mbf{h}_{q}^{s_q *} \right) ~~~ \mbox{and} ~~~ \Delta \omega_{k  p q} \equiv \omega_k^{s_k} + \omega_q^{s_q} + \omega_p^{s_p}~, 
\end{equation}
and where $k,p = \vert \bk \vert, \vert \bp \vert$.
$ \Delta \omega_{k  p q}$ is the sum of the eigen frequencies of the three modes involved in the triad. 
In general, maximum energy transfer between the three modes is ensured when the oscillations due to the detuning in the right hand side of (\ref{eq:EUleur_ODE}) are cancelled, that is, when $\Delta \omega_{k p q} \rightarrow 0$. 
If all three modes $\bk$, $\bp$ and $\bq$ are inertial waves, this leads to the well known mechanism of triadic resonance \citep{bretherton_resonant_1964,
vanneste_wave_2005,
bordes_experimental_2012}. 
When the mode $\bk$ is imposed with an amplitude $Ro$ and helicity $s_k$, $\bp$ and $\bq$ grow exponentially with a rate proportional to $k Ro$. 
This picture is changed when one of the modes, say $\bp$, is geostrophic (\ie $\omega_p^{s_p} = 0$ and $p_z = 0$, regardless of $s_p$).
The spatial interaction condition $\bk + \bp + \bq = \mbf{0}$ forces $k_z = - q_z$ and the resonance condition becomes
\begin{equation}
\label{eq:detuning}
\Delta \omega_{k p q} =  k_z \left( \frac{s_k}{k} - \frac{s_q}{q}  \right) = \omega_k \frac{1}{q} \left( q - \frac{s_q}{s_k} k  \right) \rightarrow 0~.
\end{equation} 
It may be fulfilled only when $s_q = s_k$ and $q - k \rightarrow 0$.
\textcolor{RED}{
At exact resonance, these conditions impose $\bp$ to be located on a circle centred on $-\bk_{\perp} = -\left[k_x,k_y \right]$ with radius $\vert \bk_{\perp}\vert$.
} 
\textcolor{RED}{Moreover,} in the governing equation for $\dot{b}_p^{s_p}$, the slow oscillation terms involve a coupling coefficient $C_{ p k q }^{s_p s_k s_k} \propto k - q  \rightarrow 0$. 
At exact resonance, the coupling coefficient vanishes: there is no energy transfer from waves to geostrophic modes, as proved by \cite{greenspan_non-linear_1969}. 
Nevertheless, wave-to-geostrophic transfer is still possible when the detuning $\Delta \omega_{kpq}$ is small but non-zero \citep{newell_rossby_1969,smith_transfer_1999,alexakis_rotating_2015},  and we investigate instabilities of geostrophic modes driven by this mechanism.
Let us assume that the wave $\bk$ with helicity $s_k$ is imposed with a small constant amplitude $Ro$, and that $\bp$ is geostrophic.
To infer from (\ref{eq:EUleur_ODE}) the time evolution of the geostrophic mode amplitude, we proceed to an asymptotic expansion using a two-time method involving a fast time $\tau = t$ and a slow time $T$.
\textcolor{RED}{The hierarchy between them must be a power of $kRo$, the intrinsic Rossby number based on the imposed wavelength.} 
Because the wave-to-geostrophic transfer coefficient $C_{ p k q }^{s_p s_k s_k}$ vanishes as $\Delta \omega_{kpq} \rightarrow 0$, we find via a heuristic analysis \textcolor{RED}{detailed in appendix \ref{appA}} that $T = (kRo)^2 t$, instead of $(k Ro) \,t$ for classical wave triads.
In addition, it imposes the amplitude of $\bp$ to be smaller by a factor \textcolor{RED2}{$kRo$} compared to the mode closing the triad $\bq = -\bk -\bp$.
The amplitudes of the modes interacting with the imposed waves ($\bp$ and $\bq$ with both helicity signs $s_{p,q}$) are thus expanded as
\begin{equation}
\label{eq:asymptotic_expansion_ansatz}
\left\lbrace
\begin{array}{rl}
b_q^{s_q}& = (kRo) B_{q 1}^{ s_q} (T,\tau) + (kRo)^3  \,B_{q 2}^{s_q} (T,\tau)\\
 b_p^{s_p} &=  (kRo)^2 B_{p 1}^{s_p} (T,\tau) + (kRo)^4 \, B_{p 2}^{s_p} (T,\tau)
\end{array}
 \right.
\end{equation}
where the $B_i^j$ are all $O(1)$. 
\textcolor{RED}{%
The hierarchy between orders is imposed by the need to match the slow time derivative of the leading order with the next order in the multiple scale expansion.
Since the slow  derivation introduces a factor $(kRo)^2$, there must be a $(kRo)^2$ hierarchy between the first and second orders.}
\textcolor{RED}{
To find the equations governing the leading order coeffcients, we follow the method of \cite{bretherton_resonant_1964} and inject the ansatz (\ref{eq:asymptotic_expansion_ansatz}) into (\ref{eq:EUleur_ODE}).
First, the leading order coefficients are found to be independent of $\tau$ and their long time evolution with $T$ is determined at next order.
As noted by \cite{bretherton_resonant_1964}, for fast triads with detuning larger than $O((kRo)^2)$, the imposed wave only drives fast and bounded oscillations of the second order terms, and the leading order terms must be zero to avoid secular growth.
It is only for slow triads, \ie when $\Delta \omega_{kpq} = O((kRo)^2)$, that the exponential term in equation (\ref{eq:EUleur_ODE}) drives slow oscillations that contribute to secular growth. 
Such a condition on the detuning is consistent with the numerics: the fastest growing modes at $Ro = 1 \times 10^{-2}$ ($kRo \sim 0.6$) in figure \ref{fig:growth_rate_Ro}c is $\bp = 2\pi \left[0,3,0\right]$ for which $\Delta \omega_{kpq} \simeq 0.1 \sim 0.3 (k Ro)^2$.
}
The condition $\Delta \omega_{kpq} = O((kRo)^2)$ may be fulfilled only for the modes $\bp$ with $s_p = \pm 1$ and $\bq$ with $s_q = s_k$ \textcolor{RED}{and, as noted earlier, the wave-to-geostrophic transfer coefficient $C_{ p k q }^{s_p s_k s_k}$ is also $O((kRo)^2)$.}
\textcolor{RED}{Cancelling the secular growth terms at second order gives the following amplitude equations,}
\begin{equation}
\partial_T B_{q 1}^{s_q} = \displaystyle \sum_{s_p} C_{qkp}^{s_k s_k s_p}\, B_{p_1}^{s_p \,*}\, e^{i  \textstyle \frac{\Delta \omega_{kpq}}{(kRo)^2} T  }\label{eq:amp_eq_1} ~~~
\mbox{and}~~~ \partial_T B_{p 1}^{s_p} = \frac{C_{pkq}^{s_p s_k s_k}}{(kRo)^2} \, B_{q1}^{s_q\,*}\, e^{i  \textstyle \frac{\Delta \omega_{kpq}}{(kRo)^2} T  },
\end{equation}
the rescaled quantities $C_{ p k q }^{s_p s_k s_k}/(kRo)^2$ and $\Delta \omega_{kpq}/(kRo)^2$  being $O(1)$.
These equations have exponentially growing solutions and the complex growth rate of the instability  \textcolor{RED2}{normalised by the rotation rate} is 
\begin{equation}
\label{eq:complex_growth_rate}
\sigma = i \frac{\Delta \omega_{kpq}}{2} +\frac{1}{2} \sqrt{ 4 \textstyle\sum_{s_p} C_{pkq}^{s_p s_k s_k}  C_{qkp}^{s_k s_k s_p\, *} Ro^2 - \Delta \omega_{kpq}^2 }~ \equiv i\frac{\Delta \omega_{kpq}}{2} + \sigma_k(\bp;Ro)~.
\end{equation}
Note that the product of coupling coefficients $C_{pkq}^{s_p s_k s_k}  C_{qkp}^{s_k s_k s_p\, *}$ is real. 
The expression of the real part of growth rate, $\sigma_k(\bp;Ro)$, is consistent with the numerical findings:
for a given near-resonant triad, it drops to zero at a finite value of $Ro$ and it is proportional to $Ro$ at larger $Ro$. 
However, in the small Rossby number limit, because the detuning $\Delta \omega_{kpq}$ and the coupling coefficient $C_{ p k q }^{s_p s_k s_k}$ are both $O((kRo)^2)$, the maximum geostrophic growth rate remains $O((kRo)^2)$ at most. 
The $(kRo)^2$ scaling governing the maximum growth rate of geostrophic near-resonance is found quantitatively by expanding the frequency detuning, the transfer coefficients and then $\sigma_k(\bp;Ro)$ in the neighbourhood of exact resonance. 
\textcolor{RED}{The fact that the growth rate is non-zero only close to the exact resonance is illustrated qualitatively in figure~\ref{fig:near_res_theory}a.}
%As illustrated in figure~\ref{fig:near_res_theory}a, the exactly resonant modes $\bp_0$ are located on a circle of centre $-\bk_{\perp}$ and radius \textcolor{RED}{$\vert \bk_{\perp} \vert$}, with \textcolor{RED}{$\bk_{\perp} = \left[k_x,k_y \right]$}.
%
We thus introduce $\bp = \bp_0 + (kRo)^2  \mbf{\delta}  \bp $ where $\mbf{\delta}\bp $ is a $O(1)$ vector in the geostrophic plane.
Consider the closing modes $\bq_0 = -(\bk +\bp_0)$ and $\bq = -(\bk +\bp)$, then $\bq = \bq_0 - (kRo)^2  \mbf{\delta}  \bp$.
The imposed mode $\bk$ is left unperturbed.
Since $\Delta \omega_{kpq} =C_{p_0kq_0}^{s_p s_k s_k} = 0$ at exact resonance, the leading orders of the frequency detuning and the transfer coefficients are found from (\ref{eq:detuning_coupling_definitions}) and (\ref{eq:detuning}) to be $O((kRo)^2)$. 
The perturbation of the wavevectors is thus consistent with the asymptotic expansion. 
At leading order, 
\begin{equation}
\frac{\Delta \omega_{kpq} }{(kRo)^2} \simeq s_k  \omega_k \frac{\bq_0 \cdot\mbf{\delta}  \bp }{k^2} ~~~~\mbox{and} ~~~~ \frac{C_{pkq}^{s_p s_k s_k}}{(kRo)^2} \simeq \frac{1}{2} s_k    \left(\mbf{h}_{p_0}^{s_p\,*} \cdot \left( \mbf{h}_k^{s_k\,*} \times \mbf{h}_{q_0}^{s_k\,*} \right)\right) \frac{\bq_0 \cdot\mbf{\delta}  \bp}{k},
\end{equation}
where we have used that $q_0 = k$ and $s_q = s_k$. 
In the growth rate $\sigma_k(\bp;Ro)$, the product of the coupling coefficient is
\begin{equation}
\label{eq:coupling_coefficient_sum_expansion}
\sum_{s_p} C_{pkq}^{s_p s_k s_k}  C_{qkp}^{s_k s_k s_p\, *} \simeq \frac{1}{4} (k \,Ro)^2 k^2 \frac{\bq_0 \cdot\mbf{\delta} \hat{ \bp}}{k^2}   \sum_{s_p} \left(1 - s_p \frac{p_0}{k} \right) \left| \mbf{h}_{p_0}^{s_p}\cdot \left(  \mbf{h}_k^{s_k} \times  \mbf{h}_{q_0}^{s_k} \right) \right|^2 ~. 
\end{equation}
%
%where we have used that $C_{qkp}^{s_k s_k s_p } \simeq C_{q_0 k p_0}^{s_k s_k s_p }$ at leading order. 
%
%
Therefore, at leading order in powers of $Ro$, the growth rate is
\begin{equation}
4 \sigma_k(\bp;Ro)^2 = (  \mathcal{C}_k(\bp_0)  X - \omega_k^2 X^2) (kRo)^4,
\end{equation}
where $X \equiv (\bq_0 \cdot\mbf{\delta}  \bp)/k^2$ and $\mathcal{C}_k(\bp_0)$ is the sum in the right hand side of (\ref{eq:coupling_coefficient_sum_expansion}).
When $\mathcal{C}_k(\bp_0) >0$, the growth rate reaches an optimum at $X =  \mathcal{C}_k /(2 \omega_k^2)$ with value $ (kRo)^2  \mathcal{C}_k (\bp_0) /\vert 4 \omega_k\vert $, which remains to be maximised over all exactly resonant wavevectors $ \bp_0$.
The coefficient $\mathcal{C}_k(\bp_0)$ is shown in figure \ref{fig:near_res_theory}b for several wavevectors $\bk$ with different frequencies $\omega_k$. 
When plotted against $p_0/k_{\perp}$, all the curves $\mathcal{C}_k$ collapse on a master curve that reaches a maximum value of $1$ at $p_0 \rightarrow 0$, regardless of the helicity sign $s_k$ of the imposed wave.
\textcolor{RED}{
As $(kRo) \rightarrow 0$, the  unstable geostrophic mode becomes large-scale and stems from the interaction between $\bk$ and $\bq \simeq -\bk$.}
In the low Rossby number limit, the maximal growth rate is then
\begin{equation}
\label{eq:Ro2_scaling}
\sigma_{k}^{\mathrm{max}} (Ro) =  \frac{1}{4} \frac{(k Ro)^2}{\vert \omega_k \vert}~.
\end{equation}
We \textcolor{RED}{confirm} that $k Ro$, the Rossby number based on the wavelength, is the relevant parameter to describe the growth rate of the instability. 
While each geostrophic mode follows the law (\ref{eq:complex_growth_rate}), all the growth rate curves lie below a $(k Ro)^2$ upper envelope. 
More details are given below in section \ref{sec:exact_computation} where we compare the law (\ref{eq:Ro2_scaling}) to exact computation of the growth rate curves from (\ref{eq:complex_growth_rate}).

\subsection{The moderate to large Rossby number regime }
\label{sec:large_Rossby}

The $Ro^2$ law governing the growth rate in the small Rossby number limit cannot hold as the wave amplitude is increased since the growth rate has an upper bound following a $Ro$ law.
This is proven directly by multiplying  (\ref{eq:ch2_full_problem_equation}) by $\bu$ and integrating over the fluid domain $V$, thus giving: 
\begin{equation}
\sigma =\frac{1}{2} \ddroit{ \ln \norm{\bu}_2^2 }{t} = - \frac{Ro}{\norm{u}^2_2} \int_V \bu \cdot \bnabla \bU_w \cdot \bu
\end{equation}
where $\norm{\cdot}_n$ denotes the $L_n$-norm. 
In virtue of H\"older's inequality \citep{gallet_exact_2015}, 
\begin{equation}
\label{eq:upper_bound}
\vert \sigma \vert \leq  Ro \norm{\bnabla \bU_w}_{\infty} \leq k Ro
\end{equation}
This upper bound applies in particular to the low $Ro$ scaling (\ref{eq:Ro2_scaling}), which thus holds up to $k Ro \sim 1$ at most. 
Note that, beyond that point, non-triad type instabilities (shear, centrifugal) may add to near-resonance in driving the dynamics of the flow excited by the maintained wave.

\subsection{Comparison with exact computation}
\label{sec:exact_computation}

\begin{figure}
\centering
\includegraphics[width=\linewidth]{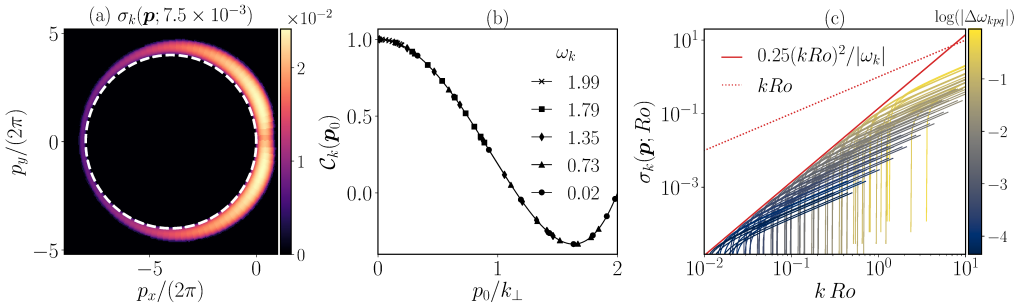}
\caption{(a) Map of the growth rate of geostrophic modes at $Ro=7.5\times 10^{-3}$ computed from (\ref{eq:complex_growth_rate}). 
The imposed wave is $2 \pi \left[ 4,0,8\right]$.
The white dashed circle locates the exactly resonant geostrophic modes.
\textcolor{RED}{The color scale gives the amplitude of the growth rate.
Where it is maximum $(p_x=0 , p_y  \simeq \pm 2)$, the detuning is about $0.04 \sim 0.24 (kRo)^2$.
}
(b) Plot of $\mathcal{C}_k (\bp_0)$ on the exact resonant circle against $p_0$ normalised by the horizontal wavenumber $k_{\perp}$ for several wavevectors $\bk$ with different frequencies $\omega_k$.
The curve is the same regardless of the imposed helicity sign $s$. 
(c) Growth rate curves of the geostrophic modes as a function of the Rossby number.
The geostrophic modes are sampled over \textcolor{RED}{15} circles whose centers are the same as the exact resonance circle with \textcolor{RED}{5} points on each circle. 
The line colour codes the frequency detuning $\vert\Delta \omega_k \vert $.
The upper envelope is compared to the law (\ref{eq:Ro2_scaling}) and the upper bound (\ref{eq:upper_bound}).
}
\label{fig:near_res_theory}
\end{figure}

%The scalings derived the the two preceding subsections are compared with the exact computation of the growth rate of geostrophic modes, given by (\ref{eq:complex_growth_rate}).
%
In figure \ref{fig:near_res_theory}, we sample the growth rate curves given by (\ref{eq:complex_growth_rate}) of many geostrophic modes in near resonant interaction with $\bk = 2 \pi \left[4,0,8\right]$, as functions of the Rossby number.
We notice that each growth rate curve follows a $Ro$ scaling at sufficiently large $Ro$, which is consistent with the asymptotic expansion carried out in section \ref{sec:single_triad}. 
However, because for all modes the growth rate vanishes at a finite value of $Ro$ (which decreases to 0 close to exact resonance), the upper envelope is a $Ro^2$ law that perfectly matches the theoretical prediction (\ref{eq:Ro2_scaling}).
This instability is thus fundamentally different from four-modes interactions for which each growth rate curve follows a  $Ro^2$ law \citep{kerswell_secondary_1999,
brunet_shortcut_2020}.
Besides, in agreement with the upper bound (\ref{eq:upper_bound}), we observe the $Ro^2$ maximum growth rate law to break down above \textcolor{RED}{$ k Ro \sim 1$}. 
Beyond that, the near-resonance growth rate derived from (\ref{eq:complex_growth_rate}) remains below the upper bound and follows a $Ro$ law.
However, additional instabilities (shear, centrifugal, etc.) may also drive the growth of geostrophic modes in this regime.

\subsection{Finite size and viscous effects}

Finite size domain translates into discretisation of the modes.
Since the exactly resonant geostrophic modes lie on a finite radius circle, they cannot be approached with arbitrarily low detuning $\Delta \omega_{kpq}$ as $Ro \rightarrow 0$.
Hence, discretisation implies the existence of a finite value of $Ro$ below which the near-resonant instability vanishes.
Regarding viscous effects, at finite but low Ekman number, \ie when $(k Ro)^2 \gg k^2 E$, the low Rossby number scaling is unaltered since the near-resonant modes $\bp$ and $\bq$ have at most similar wavenumbers to $\bk$. 
The $O(Ro)$ upper bound on the growth rate is also unaltered by the inclusion of viscosity.

\section{A refined model: the double near-resonant triad}

\begin{figure}
\centering
\includegraphics[width=0.35\linewidth]{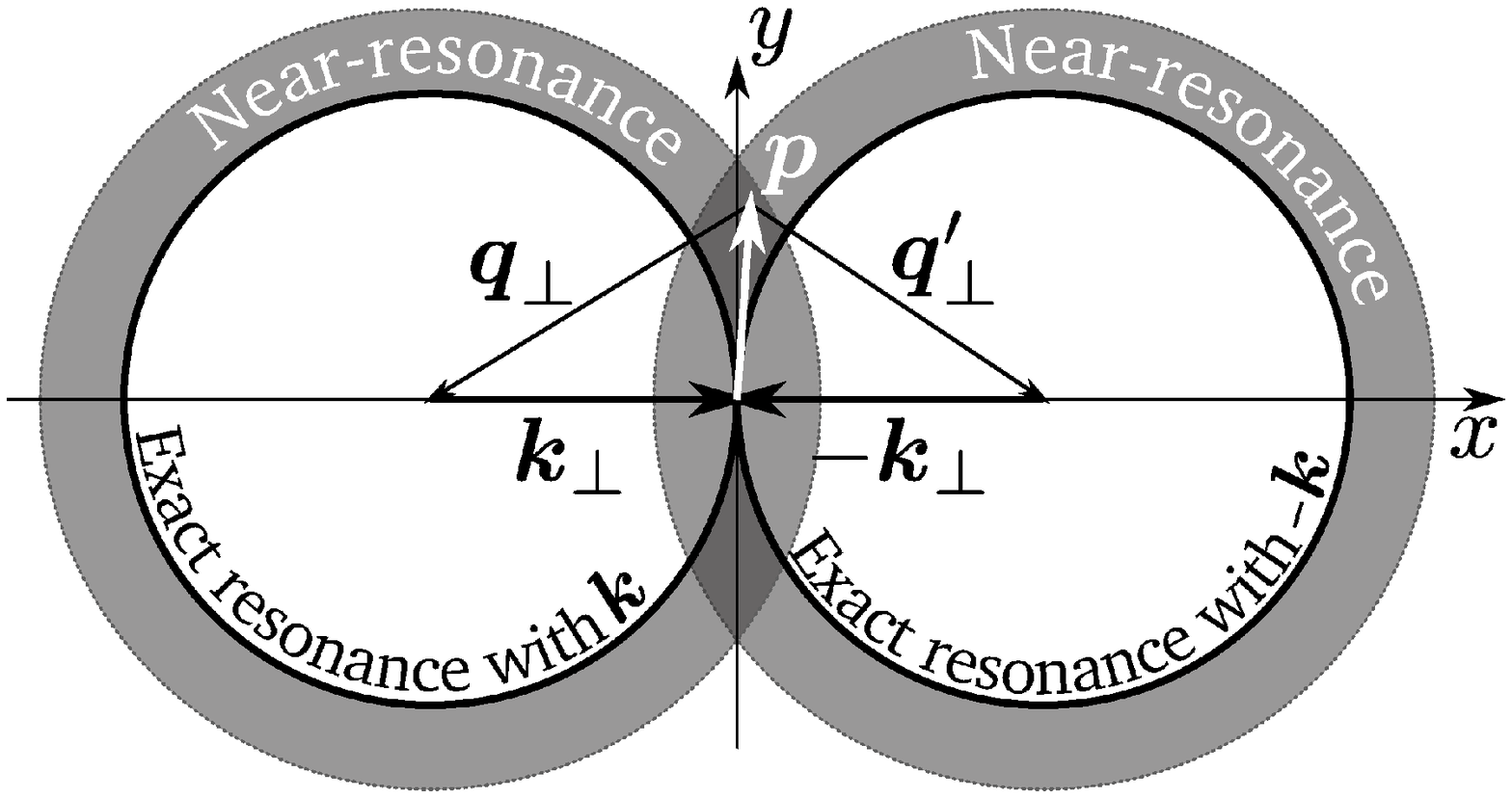}
\includegraphics[width=0.64\linewidth]{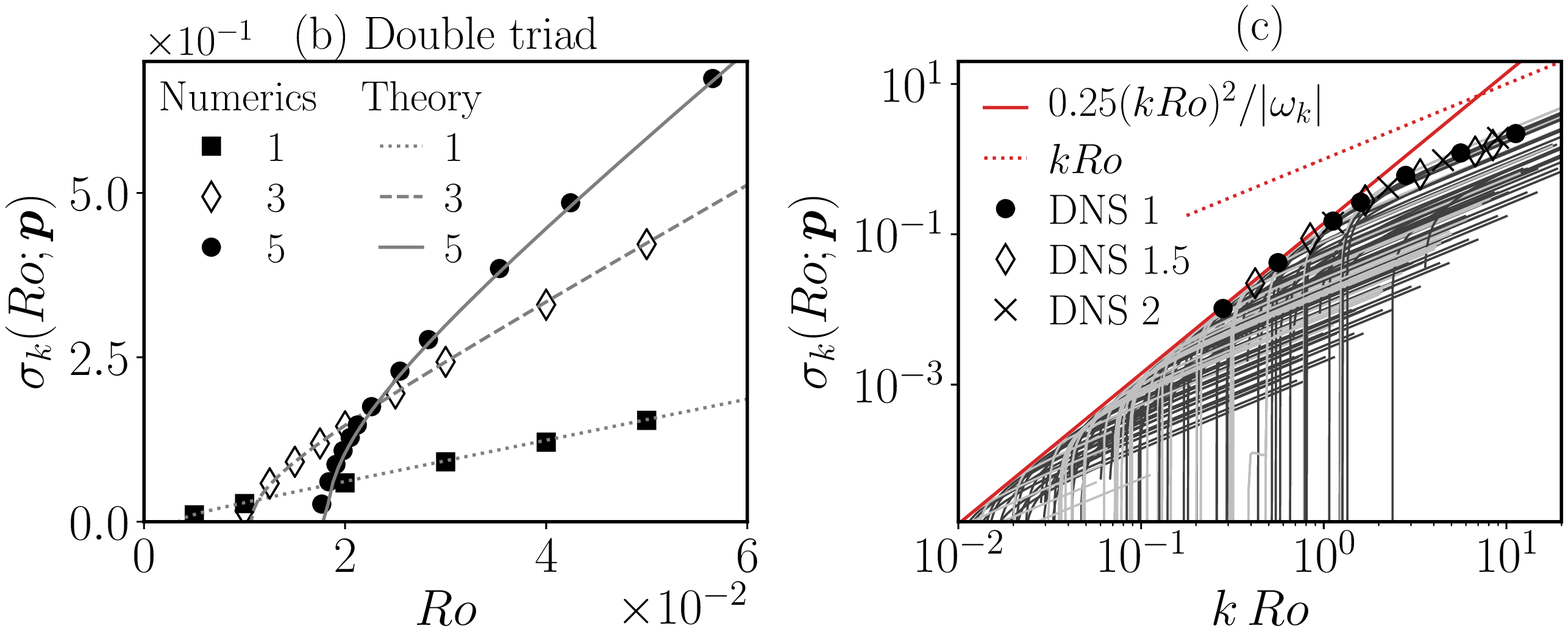}
\caption{(a) Schematic cartoon of a geostrophic mode $\bp$ in  near-resonance with both imposed modes $\pm \bk$ at the same time, based on the map of figure \ref{fig:near_res_theory}a.
The $\perp$ indices denote the horizontal component of wavevectors. 
(b) Comparison between theory (\ref{eq:growth_rate_double}) and the direct numerical simulations of figure \ref{fig:growth_rate_Ro}c. 
The imposed wave is $\bk = 2 \pi \left[4,0,8\right]$ and the geostrophic modes are $\bp = 2 \pi \left[0, p_{y}, 0\right]$, $p_{y}$ being given in the legend. 
(c) Samples of the growth rate curves $\sigma_k(\bp;Ro)$ of modes $\bp$ interacting with $\pm \bk = 2 \pi \left[4,0,8 \right]$ including simple (dark grey) and double (light grey) triad mechanisms. 
The double triad growth rate curves are determined by finding the roots of $\mathcal{P}$ (see equation (\ref{eq:characteristic_polynomial})) for wavevectors $\bp$ restricted to $\vert p_x\vert <  \pi$ and $\vert p_y \vert < 12\pi$.
We recall the $Ro^2$ law (\ref{eq:Ro2_scaling}) and the growth rate upper bound (\ref{eq:upper_bound}).
The dots represent the geostrophic growth rate found in the DNS \textcolor{RED2}{with imposed wave vector $\bk =  2 \pi \left[4,0,8 \right]$ (DNS 1), $\bk =  1.5\times 2 \pi \left[4,0,8 \right]$ (DNS 1.5) and $\bk =  2\times 2 \pi \left[4,0,8 \right]$ (DNS 2).}
}
\label{fig:refined_theory}
\end{figure}

Although promising, the model of the previous section needs to be refined to fully account for numerical results.
It misses by a factor two the growth rate curves of figure \textcolor{RED}{\ref{fig:growth_rate_Ro}c} and (\ref{eq:complex_growth_rate}) predicts a frequency $\Delta \omega_{kpq} /2 \neq 0$ for the geostrophic modes, while it is $0$ according to \textcolor{RED}{figure \ref{fig:growth_rate_Ro}d}. 
We must include in our model that not only $\bk$, but also $- \bk$ (with the same helicity sign $s$) is imposed due to Hermitian symmetry.
Consider two triads $\bk + \bp + \bq = \mbf{0}$ and $-\bk + \bp + \bq' = \mbf{0}$ (see figure \ref{fig:refined_theory}) with helicity signs $s_q = s_{q'} = s$ and $s_p = \pm 1$.
As shown schematically in figure \ref{fig:refined_theory}a, the exact resonance circles associated to $\pm \bk$ coincide at $\bp = 0$.
In a neighbourhood of this point (the dark grey intersection in figure \ref{fig:refined_theory}a), a geostrophic mode \textcolor{RED}{$\bp$} is in near-resonance with both $\pm \bk$ with both detunings $\Delta \omega_{kpq}$ and $\Delta \omega_{-kpq'}$ small.
This is all the more important that $\bp = \mbf{0}$ corresponds to the largest near-resonant growth rate as $Ro \rightarrow 0$. 
%

%
%We thus consider two triads $\bk + \bp + \bq = \mbf{0}$ and $-\bk + \bp + \bq' = \mbf{0}$ (see figure \ref{fig:refined_theory}) with helicity signs $s_q = s_{q'} = s_k$ and $s_p = \pm 1$.
%
In a very similar way to the two-time asymptotic expansion of section \ref{sec:single_triad}, we can retrieve the amplitude equations governing the slowly varying envelopes $B_{p1}^{s_p}$ and $B_{Q1}^s $, with $\bK= \pm \bk$, $\bQ = -\bK - \bp$, and $s_p = \pm 1$. 
Similarly to (\ref{eq:amp_eq_1}), the four amplitude equations are
\begin{equation}
\label{eq:full_dynamical_system}
\left\lbrace
\begin{array}{lcl}
\partial_T B_{Q1}^{s_k} &= & \displaystyle \sum_{s_p = \pm 1} C_{Q K p}^{ s_k s_k s_p} \, B_{p1}^{s_p\,*} \, e^{i \frac{\Delta \omega_{KpQ} }{ (kRo)^2} T} ,
 \\[0.4em]
\partial_T B_{p1}^{s_p} &=& \displaystyle \sum_{\bK = \pm \bk} \frac{C_{p K Q}^{ s_p s_k s_k}}{(kRo)^2} \, B_{Q1}^{s_k\,*} \, e^{i \frac{\Delta \omega_{KpQ} }{ (kRo)^2} T} .
\end{array}
\right.
\end{equation}
The characteristic polynomial $ \mathcal{P}(\sigma)$ of this set of four linear differential equations in terms of the Rossby number $Ro$ and the detunings $\Delta \omega_{K}$ is: 
\begin{equation}
\label{eq:characteristic_polynomial}
\mathcal{P}(\sigma) = \left( \sigma^2 - i \Delta \omega_{kpq} \sigma - Ro^2 S_k \right) \left( \sigma^2 - i \Delta \omega_{-kpq'} \sigma - Ro^2 S_{-k} \right)  - Ro^4 P_0~,
\end{equation}
with $S_{K}$ and $P_0$ two real coefficients defined as 
\begin{equation}
S_K = \textstyle\sum_{s_p} C_{pKQ}^{s_p s_k s_k}  C_{QKp}^{s_k s_k s_p\, *} ~\mbox{and}~ P_0  = \left(\textstyle\sum_{s_p} C_{q k p}^{ s_k s_k s_p} C_{p -k q'}^{ s_p s_k s_k} \right) \left(\textstyle\sum_{s_p} C_{q' -k p}^{ s_k s_k s_p} C_{p k q}^{ s_p s_{+k} s_k}\right).
\end{equation}
In general, the growth rate $\sigma_k(\bp;Ro)$ is found by numerical computation of the roots of $\mathcal{P}$. 
Nevertheless, an estimate of the maximum growth rate can be obtained in the limit $\vert p_x \vert \ll \vert p_y \vert$, which is relevant to our DNS.
In this case, symmetries impose $\Delta \omega_{kpq} = -\Delta \omega_{-kpq'} $, $S_k = S_{-k}$ and $P_0 = S_{k}^2$. 
The polynomial $\mathcal{P}$ then has a purely real root, 
%
%\begin{equation}
%\mathcal{P}(\sigma) = \sigma^4 - \sigma^2 \left( 2 Ro^2 C_k - \Delta \omega_k^2 \right)
%\end{equation}
%
%The non-trivial root is real, and 
%
\begin{equation}
\label{eq:growth_rate_double}
\sigma_k (\bp; Ro) = \sqrt{2 Ro^2 S_k (\bp) - \Delta \omega_{kpq}^2}~
\end{equation}
and the frequency of the growing geostrophic mode is then $0$. 
We show in figure \ref{fig:refined_theory}b the excellent agreement between this theoretical law and the DNS data of figure \ref{fig:growth_rate_Ro}a. 
%
%The root of $\mathcal{P}$ being real also explains the zero frequency of the growing geostrophic mode in figure \ref{fig:growth_rate_Ro}b.
%
An expansion similar to section \ref{sec:single_triad} reveals that the maximum growth rate follows the exact same $(kRo)^2$ law as in the case of the single triad (\ref{eq:Ro2_scaling}).
This is confirmed by the systematic computation of the growth rate in the double triad case, as shown in figure \ref{fig:refined_theory}c.
\textcolor{RED}{Note that although this instability is driven by two imposed modes, it remains different from the four-modes interaction mechanism detailed in \cite{brunet_shortcut_2020}.
The latter features intermediate, non-resonant modes that are absent in the double triad mechanism.
Moreover, the near-resonant growth rate (\ref{eq:growth_rate_double}) is proportional to $Ro$ when $Ro$ is sufficiently large, whereas it always follows a $Ro^2$ law in the case of four-modes interaction.
}

We compare in \textcolor{RED2}{figure \ref{fig:refined_theory}c} our theoretical predictions with the geostrophic growth rate extracted from DNS initiated with a large-scale noise, as in section \ref{sec:DNS_section}, down to $Ro = 5 \times 10^{-3}$.
\textcolor{RED2}{We use the same imposed wave vector as previously  ($\bk = 2 \pi \left[4,0,8 \right]$) but also 1.5 and 2 times longer wavevectors to confirm that the growth rate is a function of the intrinsic Rossby number $kRo$. 
Such a dependence is expected since in the infinite-domain limit, $L \rightarrow \infty$, $L$ becomes irrelevant and $1/k$ is the only remaining length scale, and the associated Rossby number is $k Ro$. 
}
As shown in figure \ref{fig:refined_theory}c, the numerical growth rate coincides with the maximum near-resonant growth rate in the low $Ro$ regime \textcolor{RED2}{but also in the moderate $Ro$ regime where it is proportional to $kRo$.}
Note that for the \textcolor{RED2}{three} lowest $k Ro$ points, the noise has been implemented on two-dimensional modes ($p_z = 0$) to facilitate the isolation of the geostrophic instability.
It delays the growth of two-dimensional modes with non-zero frequency by direct forcing at the lowest values of $Ro$.
Despite their rapid growth, the latter are subdominant in the saturation of wave-driven flows, and do not prevent the long-term growth of unstable geostrophic modes under the mechanism examined here.
Lastly, our analysis allows us to understand the transition in the stability of the geostrophic flow observed in the numerical study (see figure \ref{fig:amplitude_kinetic_map}).
$Ro = 2.83 \times 10^{-2}$ ($k Ro \simeq 1.6 $) corresponds to the transition zone where the geostrophic growth rate is \textcolor{RED}{$O(kRo)$}, as the wave-only triadic resonances (see figure \ref{fig:refined_theory}c).
\textcolor{RED}{When the Rossby number is decreased by one order of magnitude, as shown in figure \ref{fig:refined_theory}c, the growth rate of geostrophic modes scales like $(kRo)^2$, \ie a factor $k Ro$ smaller than the growth rate of wave-only triadic resonances.
This is why the geostrophic instability is not observed in the simulation at $Ro = 2.83 \times 10^{-3}$.
}

\vspace*{-0.25cm}

\section{Conclusion}

\textcolor{RED}{
By means of numerical simulations and theoretical analysis, we have described a new instability mechanism by which inertial waves excite $z$-invariant geostrophic modes.}
We have proved \textcolor{RED}{that this} instability is driven by near-resonant triadic interaction and derived its theoretical growth rate.
\textcolor{RED2}{
When normalised by the global rotation rate $\Omega$, the growth rate follows a $(kRo)^2$ law and a $kRo$ law at small and moderate wave amplitude, respectively, $k$ being the imposed wavenumber.
}
\textcolor{RED2}{It translates into $\Omega^{-1}$ and $\Omega^0$ laws, respectively, for the dimensional growth rate.} 
The near-resonant instability completes the picture proposed in \cite{brunet_shortcut_2020} where another inviscid geostrophic instability based on two imposed modes and four-modes interaction is detailed.
Although of different nature, both instabilities have a $(kRo)^2$ growth rate in the limit of small Rossby numbers which makes them possibly difficult to distinguish.
On the one hand, in the linear growth phase, the eigenmode of the four-mode instability consists of waves and geostrophic flow of comparable amplitude, whereas the geostrophic flow is \textcolor{RED2}{$kRo$} times smaller than the waves in the present mechanism.
On the other hand, the near-resonant instability achieves a growth rate of order \textcolor{RED2}{$kRo$ for  $kRo \gtrsim 1$}.
Our analysis may thus explain the experimental results of \cite{le_reun_experimental_2019}: they found at moderate Rossby number a geostrophic instability with a \textcolor{RED2}{growth rate proportional to $Ro$}, which thus matches the moderate \textcolor{RED2}{Rossby number} law derived here. 
\textcolor{RED}{Our work paves the way for new studies dealing with the energy transfers from waves to geostrophic modes in rotating fluids, in particular in rotating turbulence.
First, it remains to be seen under which conditions the near-resonant instability may excite slow nearly geostrophic modes with small but non-zero frequencies.
These modes have been proved to be important for the development of anisotropy in rotating turbulence \citep{smith_near_2005} even in the asymptotic limit of small Rossby numbers \citep{van_kan_critical_2019}. }
With a heuristic analysis similar to the one developed in section \ref{sec:single_triad} and appendix \ref{appA}, we can predict again a \textcolor{RED2}{$(kRo)^2$} growth rate, but a detailed analysis is difficult and remains to be done.
\textcolor{RED}{Lastly, the instability we describe in the present article could play an important role in the rotating turbulence dynamics that remains to be fully deciphered.}
Since wave-wave triadic interactions grow at rate \textcolor{RED2}{$O(kRo)$} while wave-geostrophic interactions have a \textcolor{RED2}{$O((kRo)^2)$} growth rate,
we may speculate that, at sufficiently low forcing amplitudes, rotating turbulence may remain purely three-dimensional as the wave amplitudes would remain below the threshold of geostrophic instabilities.
Such states have been observed by \cite{le_reun_inertial_2017,le_reun_experimental_2019} and \cite{brunet_shortcut_2020}, but a systematic investigation of their existence in various realisations of rotating turbulence remains to be carried out. 
\textcolor{RED}{In the future, our study could help bridging the gap between finite Rossby number experiments and simulations of turbulence \citep{godeferd_structure_2015} and asymptotic models of rotating turbulence \citep{galtier_weak_2003,bellet_wave_2006,van_kan_critical_2019}. }

\acknowledgments
\textbf{Acknowledgement.} The authors acknowledge funding by the European Research Council under the European Union's Horizon 2020 research and innovation program through Grant No. 681835-FLUDYCO-ERC-2015-CoG and FLAVE 757239. TLR is supported by the Royal Society through a Newton International Fellowship (Grant reference NIF\textbackslash R1\textbackslash 192181).

\textbf{Declaration of Interests.} The authors report no conflict of interest.

\vspace*{-0.5cm}
\appendix
\textcolor{RED}{
\section{Heuristic justification of the $(kRo)^2$ slow timescale}\label{appA}
}

To investigate near-resonant instability, we proceed to an asymptotic expansion of the spectral version of the Euler equation using a two-timing methods involving a fast time $\tau = t$ and a slow time $T$.
%
%We assume that the wave $\bk$ with helicity $s$ is imposed to a small constant amplitude $Ro$, and that $\bp$ is geostrophic.
%
The hierarchy between $\tau$ and $T$ may be derived heuristically from an analysis of the amplitude equation in a triad involving a geostrophic mode. 
Let us assume the amplitudes $b_{p,q}$ to scale like $\varepsilon_{p,q}$ and that the slow time is given by the small detuning, $ \Delta \omega_{kpq} = O(\beta)$.
We further assume that $\beta$, $\varepsilon_{p,q} = O(kRo)$ at least.
The two amplitude equations stemming from (\ref{eq:EUleur_ODE}) governing $b_{p,q}$ give the following scaling relationships:
\begin{equation}
\label{eq:amplitude hierarchy}
\beta \varepsilon_p = k Ro \beta \varepsilon_q  ~~~ \mbox{and}~~~ \beta \varepsilon_q = k Ro \varepsilon_p
\end{equation}
where we have used that the wave-to-geostrophic coupling coefficient is proportional to the detuning, hence $\beta$, as explained in section \ref{sec:single_triad} below equation (\ref{eq:detuning}). 
We have also used that the coupling coefficients scale like the wavenumber $k$.  
For the hierarchy between $\varepsilon_{p,q}$ and $\beta$ to hold, the determinant of the system (\ref{eq:amplitude hierarchy}) with unknowns $\varepsilon_{p,q}$ must vanish. This condition is satisfied when $\beta = (kRo)^2 $.
When this condition is satisfied, the equation (\ref{eq:amplitude hierarchy}) yields $\varepsilon_p = kRo \varepsilon_q$. 
Therefore, we chose the amplitudes $b_p^{s_p}$ and $b_{q}^{s_q}$ to scale like $(kRo)^2$ and $kRo$ respectively at leading order. 

\vspace*{-0.5cm}

%\bibliographystyle{jfm}
%% Note the spaces between the initials
%\bibliography{biblio}

\end{document}